\documentclass[prc,showpacs,nofootinbib,preprint]{revtex4}
\usepackage[colorlinks]{hyperref}
\usepackage{epsfig,graphics,color}
\usepackage{subfigure}
\usepackage{amssymb}
\usepackage{amsmath}

\def\pslash{p \hspace{-1.7mm}/}

\def\Dslash{D \hspace{-2.7mm}/ \;}

\def\w2{\tilde w^2}
\def\ws2{1}

\newlength{\lslash}

\def\rescale{\fontsize{8}{2}}
\begin{document}
\title{On the quark-mass dependence of the baryon ground-state masses}
\author{A. Semke and M.F.M. Lutz}
\affiliation{GSI Helmholtzzentrum f\"ur Schwerionenforschung GmbH,\\
Planckstra\ss e. 1, 64291 Darmstadt, Germany}
\date{\today}
\begin{abstract}

We perform a chiral extrapolation of the baryon octet and decuplet masses in a relativistic formulation of
chiral perturbation theory. A partial summation is assumed as implied by the use of
physical baryon and meson masses in the one-loop diagrams. Upon a  chiral expansion, our results are consistent
with strict  chiral perturbation theory at the next-to-next-to-next-to-leading order. All counter terms are
correlated by a large-$N_c$ operator analysis. Our results are confronted with recent results of unquenched
three-flavor lattice simulations. We adjust the parameter set to the pion-mass dependence of the nucleon
and omega masses as computed by the BMW Collaboration and predict the pion-mass dependence of the remaining
baryon octet and decuplet states. The current lattice simulations can be described accurately and smoothly up to
pion masses of about 600 MeV. In particular, we recover the recent results of HSC without any
further adjustments.

\end{abstract}

\maketitle

\section{Introduction}

QCD lattice simulations offer the unique opportunity to determine low-energy parameters of the
chiral Lagrangian. Currently, various lattice groups work on the baryon octet and decuplet masses in
unquenched simulations with three light quarks \cite{MILC2004,LHPC2008,PACS-CS2008,HSC2008,BMW2008,Alexandrou:2009qu}. Since
the simulations are performed also at quark masses larger than those needed to reproduce the physical hadron
masses, new information is generated that may be used to determine so-far unknown low-energy constants.

The strict chiral expansion of the baryon masses based on the flavor SU(3)
heavy-baryon formulation is poorly convergent \cite{Borasoy1997,Donoghue1998,Donoghue1999,Lehnhart2005,Semke2005,Frink2005,Frink2006,Young2009}.
This is in contrast to the two-flavor formulation, which appears to be sufficiently well converging as to justify its direct
application to lattice data \cite{Pascalutsa2005,Procura2006}. Thus, an extrapolation of the recent lattice
simulation \cite{MILC2004, LHPC2008,PACS-CS2008, HSC2008, BMW2008} down to the physical limit is not straightforward and
requires detailed studies. Indeed, at next-to-next-to-leading order (NNLO) the heavy-baryon formulation
cannot describe the latest LHP \cite{LHPC2008} and PACS-CS \cite{PACS-CS2008,PhysRevD.80.054502} lattice data.

A phenomenological remedy of the three-flavor convergence problem was suggested long ago by Donoghue and
Holstein \cite{Donoghue1998,Donoghue1999,Borasoy2002,Frink2005}. It was shown that the introduction of an
ultraviolet cutoff of the order of the kaon mass tames the large contribution of the one-loop effects.
For a recent application of such a scheme to the chiral extrapolation of the baryon masses, we refer to Ref. \cite{PhysRevD.81.014503}.
In our previous work \cite{Semke2005}, we suggested a possible alternative. A partial summation scheme was suggested where physical
meson  and baryon masses are used in the one-loop expression constituting the NNLO effects. This leads to very reasonable results
for physical quark masses. In a follow-up work \cite{Semke2007}, such a scheme was applied to a chiral extrapolation of the baryon
octet and decuplet masses. Though the overall quark-mass dependence of the early lattice results of MILC \cite{MILC2004} was
recovered roughly, the self-consistency of this approach leads to a striking and unexpected phenomenon. The system of eight
nonlinear coupled equations implies not necessarily a continuous quark-mass dependence of the baryon masses. In the more recent
work \cite{PhysRevD.82.074504}, yet a different strategy was investigated. It was shown that with a phenomenological adjustment of
the meson-baryon coupling constants in the one-loop baryon octet and decuplet self-energies evaluated at the NNLO level, one may
get close to the simulation results of PACS-CS \cite{PACS-CS2008}. In a strict chiral expansion, the effect of modifying the
meson-baryon coupling constants off their chiral SU(3) symmetric values enters at next-to-next-to-next-to-next-to-leading order
(N$^4$LO). It is yet to be clear why the N$^4$LO effects are possibly more important than the next-to-next-to-next-to-leading
order effects.

It is the purpose of the present study to extend our previous work \cite{Semke2007} and confront it with recent unquenched
lattice simulations. We will focus on the results of the BMW group \cite{BMW2008} since they provide results that can be used
directly in the continuum with negligible lattice effects. For a recent detailed study of lattice volume
effects of the results of the NPLQCD Collaboration \cite{Beane:2011pc}, we refer to Ref. \cite{Geng:2011wq}. The incorporation of the
next-to-next-to-next-to-leading order effects in our approach will allow for a quantitative extrapolation of lattice results. At this order, the meson-baryon
coupling constants are SU(3) flavor symmetric. Alltogether, there are 34 additional counter terms to be considered when going
from the NNLO  to the N$^3$LO  level. While the relevant counter terms were constructed previously in Ref. \cite{Borasoy1997} for the
baryon octet, analogous terms for the baryon decuplet will be presented fully for the first time in this work. Some terms relevant
in the decuplet sector can be found in Refs.  \cite{Jenkins1991,Banerjee1994,Jenkins1995,Pascalutsa2005,LutzSemke2010}.
Because of the large number of unknown parameters an analysis at N$^3$LO appears futile at first. However, we will demonstrate
that using large-$N_c$  correlations of the counter terms \cite{Dashen1995,Jenkins1995,Oh1999,LutzSemke2010}, a meaningful
extrapolation can be performed. The number of additional unknown  parameters is reduced significantly.

At subleading order in the large-$N_c$ expansion, we find the relevance of alltogether 20 parameters, where we find a subset
of 15 parameters to be most important. The latter are adjusted to reproduce the empirical baryon octet and decuplet masses
together with the pion-mass dependence of the nucleon and omega mass as predicted by recent QCD lattice simulations of the
BMW group \cite{BMW2008}. The smooth pion-mass dependence can be reproduced accurately. A prediction for the
pion-mass dependence of the remaining octet and decuplet masses is presented and confronted with available unquenched three-flavor simulations of various lattice groups.

\section{Chiral Lagrangian with baryon octet and decuplet fields} \label{section:chiral-lagrangian}

The construction rules for the chiral $SU(3)$ Lagrangian density are explained in Refs.  \cite{Weinberg1968,Krause1990,Ecker1989,Borasoy1997,Becher1999,Oller2006,Frink2006}.
We recall the terms relevant for our work using the notation and conventions of Refs.  \cite{Semke2005,LutzSemke2010}.
The leading order chiral Lagrangian is of the order $Q$ and is given by
\begin{eqnarray}
\mathcal{L}^{(1)} &=&
\mathrm{tr}\, \Big\{ \bar B (i\, \Dslash\, - M_{[8]})\, B \Big\}
+ F\, \mathrm{tr} \Big\{ \bar{B}\, \gamma^\mu \gamma_5\, [i\,U_\mu,B]\, \Big\} + D\, \mathrm{tr}\Big\{ \bar{B}\, \gamma^\mu \gamma_5\, \{i\,U_\mu,\,B\}\, \Big\}
\nonumber \\
&-& \mathrm{tr}\, \Big\{ \bar B_\mu \cdot \big((i\,\Dslash\, - M_{[10]})\,g^{\mu\nu} -i\,(\gamma^\mu D^\nu + \gamma^\nu D^\mu) + \gamma^\mu(i\,\Dslash + M_{[10]})\gamma^\nu \big)\, B_\nu \Big\}
\nonumber \\
&+& C\left( \mathrm{tr} \Big\{ (\bar{B}_\mu \cdot i\, U^\mu)\, B\Big\} + \mathrm{H.c.} \right)
+ H\, \mathrm{tr} \Big\{ (\bar{B}^\mu\cdot \gamma_\nu  \gamma_5\, B_\mu)\, i\,U^\nu \Big\}\,,
\label{def-L1}
\end{eqnarray}
with $U_\mu = i \,\partial_\mu \Phi/(2 f) + \cdots$.
The leading order baryon masses are given by $M_{[8]}$ and $M_{[10]}$ for the members of the flavor $SU(3)$ octet and decuplet,
respectively. The parameters $F$ and $D$ in Eq. (\ref{def-L1}) may be determined from the study of semileptonic decays of baryons,
$B\rightarrow B^\prime + e + \bar \nu_e$. This leads to  $F\simeq 0.45 $ and $D \simeq 0.80$  \cite{Okun,Lutz2002a}, the values
used in this work. The value of $C$ may be extracted from the hadronic decays of the decuplet baryons. We recall from Ref. \cite{Lutz2002a} the empirical value $C=1.6$. The parameter $H$ is poorly determined by experimental data so far \cite{Butler1993}.
Using large-$N_c$ sum rules, the parameters $C$ and $H$ may be also estimated given the empirical values for
$F$ and $D$ \cite{Dashen1994}. It holds
\begin{eqnarray}
H= 9\,F-3\,D \,,\qquad \qquad C=2\,D \,,
\label{large-Nc-HC}
\end{eqnarray}
at subleading order in the $1/N_c$ expansion.

A complete list of chiral symmetry-conserving $Q^2$ counter terms, relevant for the calculation of the N$^3$LO baryon mass
corrections, was given in Refs. \cite{Lutz2002a,LutzSemke2010}. Following these works, we group the $Q^2$ counter terms according
to their Dirac structure and display the relevant terms only. It holds:
\begin{eqnarray}\label{lbl:L_MB_four_point}
\mathcal{L}^{(2)}=\mathcal{L}^{(S)} + \mathcal{L}^{(V)}\,,
\end{eqnarray}
with
\allowdisplaybreaks[1]
\begin{eqnarray}
\mathcal{L}^{(S)} &=&- \frac{1}{2}\,g_0^{(S)}\,\mathrm{tr} \,\Big\{\bar{B}\,B \Big\}\, \mathrm{tr}\Big\{ U_\mu\, U^\mu \Big\} - \frac{1}{2}\,g_1^{(S)}\,\mathrm{tr} \,\Big\{ \bar{B}\, U^\mu \Big\}\, \mathrm{tr}\,\Big\{U_\mu\, B \Big\}
\nonumber \\
&-&\frac{1}{4}\,g_D^{(S)} \mathrm{tr}\,\Big\{\bar{B}\left\{\left\{U_\mu, U^\mu\right\}, B\right\}\Big\}
-\frac{1}{4}\,g_F^{(S)}\mathrm{tr}\,\Big\{ \bar{B}\left[\left\{U_\mu, U^\mu\right\}, B\right]\Big\}
\nonumber\\
&+& \frac{1}{2}\,h_1^{(S)}\,\mathrm{tr}\,\Big\{ \bar{B}_\mu \cdot B^\mu \Big\}\, \mathrm{tr}\,\Big\{U_\nu\; U^\nu\Big\} +
\frac{1}{2}\,h_2^{(S)}\,\mathrm{tr}\,\Big\{\bar{B}_\mu \cdot B^\nu \Big\}\, \mathrm{tr}\,\Big\{U^\mu\, U_\nu\Big\}
\nonumber \\
&+& h_3^{(S)}\,\mathrm{tr}\,\Big\{\Big(\bar{B}_\mu \cdot B^\mu\Big)\, \Big(U^\nu\, U_\nu\Big) \Big\} + \frac{1}{2}\,h_4^{(S)}\,\mathrm{tr}\,\Big\{ \Big(\bar{B}_\mu \cdot B^\nu\Big)\, \{U^\mu,\, U_\nu \} \Big\}
\nonumber \\
&+& h_5^{(S)}\, \mathrm{tr}\, \Big\{ \Big( \bar{B}_\mu \cdot U_\nu\Big)\, \Big(U^\nu\cdot B^\mu \Big) \Big\}
\nonumber \\
&+& \frac{1}{2}\,h_6^{(S)}\, \mathrm{tr} \Big\{ \Big( \bar{B}_\mu \cdot U^\mu\Big)\, \Big(U^\nu\cdot B_\nu \Big)
+\Big( \bar{B}_\mu \cdot U^\nu\Big)\, \Big(U^\mu\cdot B_\nu \Big) \Big\} \, ,
\nonumber\\
\mathcal{L}^{(V)} &=& -\frac{1}{4}\,g_0^{(V)}\, \Big( \mathrm{tr}\,\Big\{\bar{B}\, i\,\gamma^\mu\, D^\nu B\Big\} \,
\mathrm{tr}\,\Big\{ U_\nu\, U_\mu \Big\} + \mathrm{H.c.} \Big)
\nonumber \\
&-& \frac{1}{8}\,g_1^{(V)} \,\Big( \mathrm{tr}\,\Big\{\bar{B}\,U_\mu \Big\} \,i\,\gamma^\mu \, \mathrm{tr}\,\Big\{U_\nu\, D^\nu B\Big\} + \mathrm{tr}\,\Big\{\bar{B}\,U_\nu \Big\} \,i\,\gamma^\mu \, \mathrm{tr}\,\Big\{U_\mu\, D^\nu B\Big\} + \mathrm{H.c.} \Big)
\nonumber \\
&-& \frac{1}{8}\,g_D^{(V)}\, \Big(\mathrm{tr}\,\Big\{\bar{B}\, i\,\gamma^\mu \left\{\left\{U_\mu,\, U_\nu\right\}, D^\nu B\right\}\Big\} + \mathrm{H.c.} \Big)
\nonumber\\
&-& \frac{1}{8}\,g_F^{(V)}\,\Big( \mathrm{tr}\,\Big\{ \bar{B}\, i\,\gamma^\mu\, \left[\left\{U_\mu,\, U_\nu\right\},\, D^\nu B \right]\Big\} + \mathrm{H.c.} \Big)
\nonumber \\
&+& \frac{1}{4}\,h_1^{(V)}\,\Big(\mathrm{tr}\,\Big\{ \bar{B}_\lambda \cdot i\,\gamma^\mu\, D^\nu B^\lambda\Big\} \,\mathrm{tr}\,\Big\{U_\mu\, U_\nu\Big\} + \mathrm{H.c.}\Big)
\nonumber \\
&+& \frac{1}{4}\,h_2^{(V)}\,\Big(\mathrm{tr}\,\Big\{ \left(\bar{B}_\lambda \cdot i\,\gamma^\mu\, D^\nu B^\lambda \right) \{U_\mu,\, U_\nu\}\Big\} + \mathrm{H.c.}\Big)
\nonumber \\
&+& \frac{1}{4}\,h_3^{(V)}\, \Big( \mathrm{tr}\, \Big\{ \left( \bar{B}_\lambda \cdot U_\mu\right) i\,\gamma^\mu \left(U_\nu\cdot D^\nu B^\lambda \right) + \left( \bar{B}_\lambda \cdot U_\nu\right) i\,\gamma^\mu \left(U_\mu\cdot D^\nu B^\lambda \right) \Big\} + \mathrm{H.c.}\Big).\;
\label{def-Q2-terms}
\end{eqnarray}
The large number of unknown chiral parameters at this order is reduced by matching the low-energy and the $1/N_c$ expansions of the product of two axial-vector quark currents \cite{Lutz2002a,LutzSemke2010}. The 17 parameters
in Eq. (\ref{def-Q2-terms}) are correlated by the 12 sum rules
\begin{eqnarray}
&&g^{(S)}_F = g^{(S)}_0 - \frac 12\, g^{(S)}_1, \qquad h_1^{(S)}=0, \qquad h_2^{(S)}=0\,, \qquad
 h^{(S)}_3 = \frac 32\, g^{(S)}_0 - \frac 94\, g^{(S)}_1 + \frac 12\, g^{(S)}_D\,, \qquad
\nonumber \\
&& h^{(S)}_4 = 3\, \big(g^{(S)}_D+\frac 32\, g^{(S)}_1 \big)\,, \qquad
h^{(S)}_5 = g^{(S)}_D + 3\, g^{(S)}_1, \qquad h^{(S)}_6 = -3\, \big(g^{(S)}_D+\frac 32\, g^{(S)}_1 \big)\,,
\nonumber\\
&&g^{(V)}_D = -\frac 32\, g^{(V)}_1, \qquad g^{(V)}_F = g^{(V)}_0 - \frac 12\, g^{(V)}_1\,,\qquad
h^{(V)}_1 =0\,, \qquad h^{(V)}_2 = \frac 32\, g^{(V)}_0 - 3\, g^{(V)}_1\,, \qquad
\nonumber \\
&& h^{(V)}_3 = \frac 32\, g^{(V)}_1\,,
\label{result:Q2-Nc-constraints}
\end{eqnarray}
leaving only the five unknown parameters, $g_0^{(S)}, g_1^{(S)}, g_D^{(S)}$ and $g_0^{(V)}, g_1^{(V)}$.

It remains to detail the explicit symmetry-breaking terms. We collect the terms relevant
for the computation of the baryon self-energies at N$^3$LO. There are five $Q^2$ terms:
\begin{eqnarray}
&& \mathcal{L}^{(2)}_\chi = 2\, b_0 \,\mathrm{tr} \left(\bar B \,B\right) \mathrm{tr}\left(\chi_+\right) + 2 \,b_D\,\mathrm{tr}\left(\bar{B}\,\{\chi_+,\,B\}\right) + 2\, b_F\,\mathrm{tr}\left(\bar{B}\,[\chi_+,B]\right) \nonumber \\
&& \qquad  -\, 2\, d_0\, \mathrm{tr}\left(\bar B_\mu \cdot B^\mu\right) \mathrm{tr}(\chi_+) - 2\, d_D\, \mathrm{tr} \left( \left(\bar{B}_\mu \cdot B^\mu\right) \chi_+\right)\,,
\nonumber\\
&& \chi_+ = \chi_0 - \frac{1}{8\,f^2}\,\big\{ \Phi, \big\{ \Phi, \,\chi_0 \big\} + {\mathcal O} \left( \Phi^4 \right) \,,
\label{def-chi-2}
\end{eqnarray}
with $\chi_0 = 2\,B_0 \,{\rm diag }(m ,m, m_s)  $ proportional to the quark-mass matrix. We do not consider
isospin-violating effects in this work, and we use $f= 92.4$ MeV. A matching of the chiral interaction terms (\ref{def-chi-2})
to the large-$N_c$ operator analysis for the baryon masses in Refs. \cite{Dashen1995,Jenkins1995} leads to one sum rule:
\begin{eqnarray}
b_D + b_F= \frac 13 \,d_D\,,
\label{largeNcbd}
\end{eqnarray}
accurate to subleading order in the $1/N_c$ expansion.

We continue with the symmetry-breaking part of the chiral Lagrangian.
There are  five $Q^3$ terms and 12 $Q^4$ terms:
\allowdisplaybreaks[1]
\begin{eqnarray}
\mathcal{L}^{(3)}_\chi &=& \zeta_0\, \mathrm{tr} \big(\bar{B}\, (i\,\Dslash -M_{[8]})\, B\big)\, \mathrm{tr}(\chi_+) + \zeta_D\, \mathrm{tr} \big(\bar{B}\, (i\,\Dslash -M_{[8]})\, [B, \chi_+]\big) \nonumber\\
&+& \zeta_F\, \mathrm{tr} \big(\bar{B}\, (i\,\Dslash -M_{[8]})\, \{B, \chi_+\} \big) \nonumber \\
&-& \xi^{}_0\, \mathrm{tr} \big(\bar B_\mu\, (i\,\Dslash -M_{[10]})\, B^\mu \big)\, \mathrm{tr}(\chi_+) - \xi^{}_D\, \mathrm{tr} \big(\bar B_\mu\, (i\,\Dslash -M_{[10]})\,B^\mu\, \chi_+ \big) \,,
\nonumber\\
\mathcal{L}^{(4)}_\chi &=& c_0\, \mathrm{tr}\left(\bar B\, B\right) \mathrm{tr} \left(\chi_+^2\right) + c_1\, \mathrm{tr} \left(\bar B \,\chi_+\right) \mathrm{tr}\left(\chi_+ B\right) \nonumber \\
&+& c_2\, \mathrm{tr} \left( \bar B\, \{\chi_+^2,\, B\} \right)  + c_3\,\mathrm{tr} \left( \bar B\, [\chi_+^2, \,B] \right)
\nonumber \\
&+& c_4\, \mathrm{tr} \left(\bar B\, \{\chi_+,\,B\} \right) \mathrm{tr} (\chi_+) + c_5\, \mathrm{tr}\left(\bar B\, [\chi_+,\,B]\right) \mathrm{tr} (\chi_+) \nonumber \\
&+& c_6\, \mathrm{tr} \left(\bar B \,B\right) \left(\mathrm{tr}(\chi_+)\right)^2
\nonumber \\
&-& e_0\, \mathrm{tr}\left(\bar B_\mu \cdot B^\mu \right) \mathrm{tr}\left(\chi_+^2\right) - e_1\, \mathrm{tr}\left( \left(\bar{B}_\mu \cdot \chi_+\right) \left(\chi_+ \cdot B^\mu\right) \right)
\nonumber \\
&-& e_2\, \mathrm{tr}\left( \left(\bar B_\mu \cdot B^\mu\right)\cdot \chi_+^2\right) - e_3\, \mathrm{tr}\left( \left(\bar B_\mu \cdot B^\mu\right)\cdot \chi_+\right) \mathrm{tr}(\chi_+) \nonumber \\
&-& e_4\, \mathrm{tr}\left(\bar B_\mu \cdot B^\mu\right) \left(\mathrm{tr}(\chi_+)\right)^2\,.
\label{def-chi-34}
\end{eqnarray}
We consider again large-$N_c$ sum rules for the parameters introduced in Eq. (\ref{def-chi-34}). For $\zeta_0, \zeta_D, \zeta_F$, and
$\xi_0, \xi_D$, the sum rules are analogous to the ones for the $Q^2$ terms, i.e. it holds:
\begin{eqnarray}
\zeta_D + \zeta_F= \frac 13 \,\xi^{}_D\,,
\label{result:large-Nc-zeta}
\end{eqnarray}
at subleading order. A matching of the chiral interaction terms (\ref{def-chi-34}) to the
large-$N_c$ operator analysis for the baryon masses in Ref. \cite{Jenkins1995} leads to the seven sum rules:
\begin{gather}
c_0 = \frac{1}{2} \,c_1, \qquad c_2=-\frac{3}{2} \,c_1,\qquad c_3=0 \,, \qquad
\nonumber\\
e_0 = 0\,, \qquad e_1 = -2 \,c_2, \qquad  e_2 = 3\, c_2\,, \qquad e_3 = 3\, (c_4 + c_5)\,,
\label{result:large-Nc-chi}
\end{gather}
valid at NNLO in the expansion. Assuming the approximate validity of Eq. (\ref{result:large-Nc-chi}), it suffices
to determine the five parameters $c_1,c_4,c_5, c_6$, and $e_4$.

The terms in $\mathcal{L}^{(3)}_\chi$ are redundant upon a suitable redefinition of the baryon fields.
Why do we consider such terms at all? The reason is that the redundance of the
parameters $\zeta_0,\zeta_D, \zeta_F$, and $\xi_0, \xi_D$ is lifted once we insist on the
large-$N_c$ relations (\ref{result:large-Nc-chi}). This is seen by eliminating $\mathcal{L}^{(3)}_\chi$ in application of the equation of motion for the baryon fields. A renormalization of terms
already  present in the chiral Lagrangian, in particular, those in  $\mathcal{L}^{(4)}_\chi$, arises. We find the renormalized or effective coupling strengths:
\begin{align}
c^{\rm eff}_0 &=  - 2 \left(\zeta_D\, b_D - \zeta_F\, b_F \right) + c_0, & e^{\rm eff}_0 &=
e_0, \nonumber \\
c^{\rm eff}_1 &=  -{\textstyle{ 4\over 3}} \left(\zeta_D\, b_D - 3 \,\zeta_F\, b_F \right) + c_1, & e^{\rm eff}_1
&=  {\textstyle{ 4\over 3}} \, \xi_D\, d_D + e_1, \nonumber \\
c^{\rm eff}_2 &=  2 \left(\zeta_D\, b_D - 3 \,\zeta_F\, b_F \right) + c_2, & e^{\rm eff}_2 &=
-2\,\xi_D \,d_D + e_2, \nonumber \\
c^{\rm eff}_3 &= -2 \left(\zeta_D\, b_F + \zeta_F\, b_D \right) + c_3, & e^{\rm eff}_3 &= - 2\left(
\xi_0 \,d_D + \xi_D\, d_0 \right) + e_3, \nonumber \\
c^{\rm eff}_4 &= -2 \left(\zeta_0\, b_D + \zeta_D \,b_0 + 2\,\zeta_D\, b_D - 2\,\zeta_F\, b_F \right)
+ c_4, & e^{\rm eff}_4 &= -2\,\xi_0 \,d_0 + e_4, \nonumber\\
c^{\rm eff}_5 &= -2 \left(\zeta_0 \,b_F + \zeta_F\, b_0 \right) + c_5, && \nonumber \\
c^{\rm eff}_6 &= -2 \left(\zeta_0 \,b_0 - \zeta_D \,b_D + \zeta_F\, b_F \right) + c_6. &&
\label{def-decomposition}
\end{align}
Inserting Eq. (\ref{def-decomposition}) into the sum  rules (\ref{result:large-Nc-chi}) reveals that there
is no nontrivial way to dial the parameters $\zeta_D, \zeta_F$, and $\xi_D$ as to be compatible with Eq.  (\ref{result:large-Nc-chi}). It follows that
those parameters are independent of the five parameters $c_1,c_4,c_5, c_6$ and $e_4$. For the singlet parameters one finds the correlation
\begin{eqnarray}
3\,\zeta_0\,(b_D+b_F) = \xi_0 \,d_D \,,
\label{res-zetas}
\end{eqnarray}
which is consistent with the trivial solution $\zeta_0 = 0 = \xi_0$. The results (\ref{def-decomposition}, \ref{res-zetas})
illustrate that the N$^3$LO effect of a variation of the parameters $\zeta_0 $ and $\xi_0$ as correlated by Eq. (\ref{res-zetas})
can be reproduced by a suitable variation of the parameters $c_{4,5,6}$ and $e_4$. We conclude that given the sum rules, 
(\ref{result:large-Nc-zeta}, \ref{result:large-Nc-chi})
there are alltogether 8 symmetry-breaking parameters at N$^3$LO: the parameters $c_1,c_4,c_5, c_6, e_4$, and $\zeta_F, \xi_D$
together with either $\xi_0$ or $\zeta_0$.

In anticipation of our results, we state the crucial importance of the parameters $\zeta_F, \xi_D$. Within the self-consistent
approach applied in this work, they lead to a smooth chiral extrapolation of the baryon masses. Any attempt with $\zeta_F= \xi_D=0$
to establish a smooth chiral extrapolation would be futile.

\section{Chiral loop expansion of the baryon masses}
We turn to the computation of the baryon masses. The baryon self-energy, $\Sigma_B(\pslash)$, may be considered to be
a function of $p_\mu \gamma^\mu $ only, with the 4-momentum $p_\mu$ of the baryon $B$. This is obvious for the spin one-half
baryons, but less immediate for the spin-three-half baryons. We refer to Ref. \cite{Semke2005} for technical details.
To order $Q^4$, the self-energy receives contributions from tree-level diagrams and one-loop diagrams:
\begin{eqnarray}
\Sigma_B(M_B) =  \Sigma^{\rm tree-level}_B + \Sigma^{\rm loop }_B \,,
\label{def-Sigma}
\end{eqnarray}
where the index $B$ stands for the members of the flavor $SU(3)$ octet and decuplet, $B\in[8], [10]$.
The separation of the baryon self-energies into a loop and a tree-level
contribution is not unique depending on the renormalization scheme. In this work, we apply the $\chi$MS scheme developed in Ref. \cite{Semke2005}, where we keep the dependence on the ultraviolet renormalization scale only. A possible dependence on an infrared
renormalization scale is ignored in this work as to be close to more conventional renormalization schemes.
A matching with alternative renormalization schemes is most economically performed by a direct comparison
with the explicit expressions of this section. Our renormalized tree-level self-energies are collected in the Appendix.

The physical mass of the baryon $M_B$ is determined by the condition
\begin{eqnarray}
M_B -\Sigma_B(M_B)    = \left\{\begin{array}{ll}
\bar M_{[8]}  & \qquad {\rm for } \qquad B\in[8]\\
\bar M_{[10]}  & \qquad {\rm for }\qquad B \in[10]
\end{array} \right. \,,
\label{def-non-linear-system}
\end{eqnarray}
where $\bar M_{[8]} $ and $\bar M_{[10]} $ are the renormalized and scale-independent bare masses of the baryon octet and decuplet.
We consistently use a bar for renormalized quantities throughout this work. At NNLO, the one-loop contributions probe the
coupling constants $F,D, C$, and $H$ introduced in Eq. (\ref{def-L1}).
Complete expressions for baryon octet and decuplet states were first established in Ref. \cite{Semke2005}. Partial
results are documented in Ref. \cite{Banerjee1995}. Here, we complement our previous result by additional contributions
from the counter terms (\ref{def-Q2-terms}, \ref{def-chi-34}), which turn relevant at N$^3$LO. For previous
N$^3$LO studies in the baryon octet sector see Refs. \cite{Borasoy1997,Frink2004}. Some partial results in the baryon decuplet
sector can be found in \cite{Tiburzi2004}. Alltogether, we obtain for the renormalized loop contribution the expressions
\allowdisplaybreaks[1]
\begin{eqnarray}
&&\Sigma^{\rm loop}_{B \in [8]} = \sum_{Q\in [8], R\in [8]}
\left(\frac{G_{QR}^{(B)}}{2\,f} \right)^2  \Bigg\{
- \frac{(M_B+M_R)^2}{E_R+M_R}\, p^2_{QR}\,
\Big(\bar I_{QR} + \frac{\bar I_Q}{M_R^2-m_Q^2}\,\Big)
+\frac{M_R^2-M_B^2}{2\,M_B}\, \bar I_Q
\Bigg\}
\nonumber \\
&& \qquad  \;\,\,\,+\sum_{Q\in [8], R\in [10]}
\left(\frac{G_{QR}^{(B)}}{2\,f} \right)^2 \, \Bigg\{
 - \frac{2}{3}\,\frac{M_B^2}{M_R^2}\,\big(E_R+M_R\big)\,p_{QR}^{\,2}\,
\Big(\bar I_{QR} + \frac{\bar I_Q}{M_R^2-m_Q^2}\Big)
\nonumber\\
&& \qquad \qquad
+ \,\Bigg(  \frac{(M_R-M_B)\,(M_R+M_B)^3+m_Q^4}{12\,M_B\,M^2_R}\,
+ \frac{5\,M_B^2+6\,M_R\,M_B-2\,M_R^2}{12\,M_B\,M_R^2}\,m_Q^2\Bigg)\,\bar I_Q
   \Bigg\}
\nonumber\\
&& \qquad  \;\,\,\,+ \,
\frac{1}{(2\,f)^2}\sum_{Q\in [8]} \Bigg(  G^{(\chi )}_{BQ} - m_Q^2\,G^{(S)}_{BQ} - \frac 14 \, m_Q^2\,M_B \,G^{(V)}_{BQ}\Bigg)\, \bar I_Q \,,
\label{result-loop-8}
\end{eqnarray}
and
\allowdisplaybreaks[1]
\begin{eqnarray}
&&\Sigma^{\rm loop}_{B\in [10]} = \sum_{Q\in [8], R\in [8]}
\left(\frac{G_{QR}^{(B)}}{2\,f} \right)^2  \Bigg\{
-\frac{1}{3}\,\big( E_R +M_R\big)\,p_{QR}^{\,2}\,
\Big(\bar I_{QR}+ \frac{\bar I_Q}{M_R^2-m_Q^2}\Big)
\nonumber\\
&& \qquad \qquad
+ \,\Bigg( \frac{(M_R-M_B)\,(M_R+M_B)^3 + m_Q^4}{24\,M^3_B}
- \frac{3\,M_B^2+2\,M_R\,M_B+2\,M_R^2}{24\,M^3_B}\,m_Q^2\Bigg)\,\bar I_Q \Bigg\}
\nonumber\\
&& \qquad \;\,\,\,+\sum_{Q\in [8], R\in [10]}
\left(\frac{G_{QR}^{(B)}}{2\,f} \right)^2 \, \Bigg\{
 -\frac{(M_B+M_R)^2}{9\,M_R^2}\,\frac{2\,E_R\,(E_R-M_R)+5\,M_R^2}{E_R+M_R}\,
p_{QR}^{\,2}\,\Big(\bar I_{QR}(M_B^2)
\nonumber\\
&&\qquad \qquad
+ \,\frac{\bar I_Q}{M_R^2-m_Q^2}\Big)
+\,\Bigg( \frac{M_R^4+M_B^4+12\,M_R^2\,M_B^2-2\,M_R\,M_B\,(M_B^2+M_R^2)}{36\,M^3_B\,M^2_R}\,
(M^2_R-M^2_B)
\nonumber\\
&&\qquad \qquad  +\,\frac{(M_B+M_R)^2\,m_Q^4}{36\,M_B^3\,M_R^2}
+\frac{3\,M_B^4-2\,M^3_B\,M_R+3\,M_B^2\,M_R^2-2\,M_R^4}{36\,M_B^3\,M_R^2}
\,m_Q^2\Bigg)\,\bar I_Q \Bigg\}
\nonumber\\
&& \qquad  \;\,\,\,+\,
\frac{1}{(2\,f)^2}\sum_{Q\in [8]} \Bigg(  G^{(\chi )}_{BQ} - m_Q^2\,G^{(S)}_{BQ} - \frac 14 \, m_Q^2\,M_B \,G^{(V)}_{BQ}\Bigg)\, \bar I_Q \,,
\label{result-loop-10}
\end{eqnarray}

\begin{table}
\rescale
\setlength{\tabcolsep}{1.mm}
\setlength{\arraycolsep}{0.2mm}
\renewcommand{\arraystretch}{1.}
\begin{center}
 \begin{tabular}{|c c|c|c|c|}\hline
$B$ & $Q$ & $G^{(\chi)}_{BQ}$ & $G^{(S)}_{BQ}$ & $G^{(V)}_{BQ}$ \\ \hline

    & $\pi$ & $ 24\,B_0\,m\, (2\,\bar b_0 +\bar b_D +\bar b_F)$ & $3\,g_0^{(S)} +\frac{3}{2} \,g_D^{(S)} +\frac{3}{2}\,  g_F^{(S)}$ & $ 3\,g_0^{(V)} +\frac{3}{2}\, g_D^{(V)} +\frac{3}{2}\, g_F^{(V)}  $\\
$N$ & $K$ & $8\,B_0\,(m+m_s)\,(4\,\bar b_0+3\,\bar b_D-\bar b_F)$ & $4\,g_0^{(S)} + g_1^{(S)} + 3\,g_D^{(S)} - g_F^{(S)}$ & $4\,g_0^{(V)} + g_1^{(V)} + 3\,g_D^{(V)} - g_F^{(V)}$  \\
    & $\eta$ & $\begin{array}{l} \frac{8}{3}\,B_0\,m\,(2\,\bar b_0+\bar b_D+\bar b_F)  \\ +\frac{32}{3}\,B_0\,m_s\,(\bar b_0+\bar b_D-\bar b_F)  \end{array} $
     & $ g_0^{(S)} + \frac{5}{6}\,g_D^{(S)} - \frac{1}{2}\,g_F^{(S)} $ & $ g_0^{(V)} + \frac{5}{6}\,g_D^{(V)} - \frac{1}{2}\,g_F^{(V)}$ \\ \hline

    &  $\pi$ &$16\,B_0\,m\, (3\,\bar b_0 +\bar b_D)$  & $3\,g_0^{(S)} + g_D^{(S)}$ & $3\,g_0^{(V)} + g_D^{(V)}$ \\
$\Lambda$ & $K$ & $\frac{16}{3}\,B_0\,(m+m_s)\,(6\,\bar b_0+5\,\bar b_D)$ & $ 4\,g_0^{(S)} + \frac{10}{3}\,g _D^{(S)} $ & $ 4\,g_0^{(V)} + \frac{10}{3} \,g _D^{(V)} $ \\
    & $\eta$ & $\begin{array}{l} \frac{16}{9}\,B_0\,m\,(3\,\bar b_0+\bar b_D)  \\ +\frac{32}{9}\,B_0\,m_s\,(3\,\bar b_0+4\,\bar b_D)  \end{array} $
    & $g_0^{(S)} + g_1^{(S)} + g_D^{(S)}$ & $g_0^{(V)} + g_1^{(V)} + g_D^{(V)}$ \\ \hline

    & $\pi$ & $48\,B_0\,m\, (\bar b_0 +\bar b_D)$ & $3\,g_0^{(S)} + g_1^{(S)} + 3\,g_D^{(S)}$ & $3\,g_0^{(V)} + g_1^{(V)} + 3\,g_D^{(V)}$ \\
$\Sigma$ & $K$ & $16\,B_0\,(m+m_s)\,(2\,\bar b_0+\bar b_D)$ & $4\,g_0^{(S)} + 2\,g_D^{(S)} $  & $4\,g_0^{(V)} + 2\,g_D^{(V)} $ \\
    & $\eta$ & $\frac{16}{3}\,B_0 \left( m\,(\bar b_0+\bar b_D) +2\,m_s\,\bar b_0 \right)$ & $ g_0^{(S)} + \frac{1}{3}\,g_D^{(S)} $ & $ g_0^{(V)} + \frac{1}{3}\,g_D^{(V)} $ \\ \hline

    & $\pi$ &$24\,B_0\,m\, (2\,\bar b_0 +\bar b_D -\bar b_F)$ & $3\,g_0^{(S)} + \frac{3}{2} \,g_D^{(S)} - \frac{3}{2} \,g_F^{(S)} $ & $ 3\,g_0^{(V)} + \frac{3}{2} \,g_D^{(V)} - \frac{3}{2} \,g_F^{(V)} $ \\
$\Xi$ &  $K$ &$8\,B_0\,(m+m_s)\,(4\,\bar b_0+3\,\bar b_D+\bar b_F)$ & $4\,g_0^{(S)} + g_1^{(S)} + 3\,g_D^{(S)} + g_F^{(S)} $ & $4\,g_0^{(V)} + g_1^{(V)} + 3\,g_D^{(V)} + g_F^{(V)}$  \\
    & $\eta$  & $\begin{array}{l} \frac{8}{3}\,B_0\,m\,(2\,\bar b_0+\bar b_D-\bar b_F)  \\ +\frac{32}{3}\,B_0\,m_s\,(\bar b_0+\bar b_D+\bar b_F)  \end{array} $
    & $g_0^{(S)} + \frac{5}{6}\, \,g_D^{(S)} + \frac{1}{2}\,g_F^{(S)} $ & $ g_0^{(V)} + \frac{5}{6}\,g_D^{(V)} + \frac{1}{2}\,g_F^{(V)} $ \\ \hline \hline

    & $\pi$ & $24\,B_0\,m\, (2\,\bar d_0 +\bar d_D)$ &$ 3\,\tilde h_1^{(S)} + 3\,\tilde h_2^{(S)} + 2\,\tilde h_3^{(S)}$ & $ 3\,h_1^{(V)} + 3\,h_2^{(V)} + 2\,h_3^{(V)}$\\
$\Delta$ & $K$ & $8\, B_0\,(m+m_s)\,(4\,\bar d_0+\bar d_D)$ & $ 4\,\tilde h_1^{(S)} + 2\,\tilde h_2^{(S)} + 2\,\tilde h_3^{(S)}  $ &$ 4\,h_1^{(V)} + 2\,h_2^{(V)} + 2\,h_3^{(V)}  $ \\
    & $\eta$ & $ \frac{8}{3}\,B_0\,( m\,(2\,\bar d_0+\bar d_D) + 4\,m_s\,\bar d_0)  $
    &$ \tilde h_1^{(S)} + \frac{1}{3}\,\tilde h_2^{(S)} $ & $ h_1^{(V)} + \frac{1}{3} \,h_2^{(V)} $\\ \hline

    & $\pi$ & $16\,B_0\,m\, (3\,\bar d_0 +\bar d_D)$ &$3\,\tilde h_1^{(S)} + 2\,\tilde h_2^{(S)} + \frac{5}{3}\,\tilde h_3^{(S)}$ & $3\,h_1^{(V)} + 2\,h_2^{(V)} + \frac{5}{3}\,h_3^{(V)}$\\
$\Sigma^*$ & $K$ & $\frac{32}{3}\,B_0\,(m+m_s)\,(3\,\bar d_0+\bar d_D)$  & $ 4\,\tilde h_1^{(S)} + \frac{8}{3}\,\tilde h_2^{(S)} + \frac{4}{3}\,\tilde h_3^{(S)} $ & $ 4\,h_1^{(V)} + \frac{8}{3}\,h_2^{(V)} + \frac{4}{3}\,h_3^{(V)} $ \\
    & $\eta$ & $\frac{16}{9}\,B_0\, (m+2\,m_s)(3\,\bar d_0+\bar d_D)$& $\tilde h_1^{(S)} + \frac{2}{3}\,\tilde h_2^{(S)} + \tilde h_3^{(S)}$ & $h_1^{(V)} + \frac{2}{3}\,h_2^{(V)} + h_3^{(V)}$\\ \hline

    & $\pi$ & $8\,B_0\,m\, (6\,\bar d_0 +\bar d_D)$&$3\,\tilde h_1^{(S)} + \tilde h_2^{(S)} + \tilde h_3^{(S)}$ & $3\,h_1^{(V)} + h_2^{(V)} + h_3^{(V)}$\\
$\Xi^*$ &  $K$ &$\frac{8}{3}\,B_0\,(m+m_s)\,(12\,\bar d_0+5\,\bar d_D)$ & $4\,\tilde h_1^{(S)} + \frac{10}{3}\,\tilde h_2^{(S)} + 2\,\tilde h_3^{(S)}$ & $4\,h_1^{(V)} + \frac{10}{3}\,h_2^{(V)} + 2\,h_3^{(V)}$  \\
    & $\eta$ & $\begin{array}{l} \frac{8}{9}\,B_0\,m\,(6\,\bar d_0+\bar d_D)  \\ +\frac{8}{9}\,B_0\,m_s\,(12\,\bar d_0 +8\,\bar d_D)  \end{array} $
     & $\tilde h_1^{(S)} + \tilde h_2^{(S)} + \tilde h_3^{(S)}$ & $h_1^{(V)} + h_2^{(V)} + h_3^{(V)}$\\ \hline

    & $\pi$ &$48\,B_0\,m\,\bar d_0$ & $3\,\tilde h_1^{(S)}$ & $3\,h_1^{(V)}$\\
$\Omega$ & $K$ &$16\,B_0\,(m+m_s)\,(2\,\bar d_0+\bar d_D)$ & $4\,\tilde h_1^{(S)} + 4\,\tilde h_2^{(S)} + 4\,\tilde h_3^{(S)} $ & $4\,h_1^{(V)} + 4\,h_2^{(V)} + 4\,h_3^{(V)} $  \\
    & $\eta$ & $\frac{16}{3}\,B_0\,(m\,\bar d_0 +2\,m_s(\bar d_0+\bar d_D))  $
    & $\tilde h_1^{(S)} + \frac{4}{3}\,\tilde h_2^{(S)}$ & $h_1^{(V)} + \frac{4}{3}\,h_2^{(V)}$\\ \hline

 \end{tabular}
\caption{Coefficients $G^{(\chi)}_{BQ}$, $G^{(S)}_{BQ}$ and $G^{(V)}_{BQ}$. }
\label{tab:G-chi}
\end{center}
\end{table}
where
\begin{eqnarray}
&& \bar I_Q =\frac{m_Q^2}{(4\,\pi)^2}\,
\ln \left( \frac{m_Q^2}{\mu_{\,UV}^2}\right)\,,
\nonumber\\
&& \bar I_{Q R}=\frac{1}{16\,\pi^2}
\left\{ \left(\frac{1}{2}\,\frac{m_Q^2+M_R^2}{m_Q^2-M_R^2}
-\frac{m_Q^2-M_R^2}{2\,M_B^2}
\right)
\,\ln \left( \frac{m_Q^2}{M_R^2}\right)
\right.
\nonumber\\
&& \;\quad \;\,+\left.
\frac{p_{Q R}}{M_B}\,
\left( \ln \left(1-\frac{M_B^2-2\,p_{Q R}\,M_B}{m_Q^2+M_R^2} \right)
-\ln \left(1-\frac{M_B^2+2\,p_{Q R}\,M_B}{m_Q^2+M_R^2} \right)\right)
\right\}\;,
\nonumber\\
&& p_{Q R}^2 =
\frac{M_B^2}{4}-\frac{M_R^2+m_Q^2}{2}+\frac{(M_R^2-m_Q^2)^2}{4\,M_B^2} \,,\qquad
E_R^2=M_R^2+p_{QR}^2 \,.
\label{def-tadpole}
\end{eqnarray}
The sums in Eqs. (\ref{result-loop-8}, \ref{result-loop-10}) extend over the intermediate Goldstone bosons ($Q\in[8]$) baryon
octet ($R\in [8]$) and decuplet  states ($R\in[10]$). The coupling constants $G_{QR}^{(B)}$  are
determined by the parameters $F,D,C,H$. They are listed in Ref. \cite{Semke2005}. The coupling constants
$G_{QR}^{(\chi)}$ probe the renormalized symmetry-breaking parameters $\bar b_0,\bar b_D, \bar b_F, \bar d_0$, and $\bar d_D$. They are detailed
in Table \ref{tab:G-chi}, together with $G_{QR}^{(S)}$ and $G_{QR}^{(V)}$ which are proportional to the
symmetry-preserving parameters introduced in Eq. (\ref{def-Q2-terms}). In Table \ref{tab:G-chi}, we apply the notation
\begin{eqnarray}
\tilde h_1^{(S)}\equiv h_1^{(S)} + \frac{1}{4}\,h_2^{(S)}\,,\qquad
\tilde h_2^{(S)}\equiv h_3^{(S)} + \frac 14\, h_4^{(S)}\,, \qquad
\tilde h_3^{(S)}\equiv h_5^{(S)} + \frac 14 \,h_6^{(S)} \,.
\label{def-tilde-h}
\end{eqnarray}
The 6 parameters $h_{1-6}^{(S)}$  enter the decuplet self-energy in the three combinations (\ref{def-tilde-h}) only.

\begin{table}[t]
\begin{center}
\begin{tabular}{|c|c|}
\hline
&Empirical value at  $\mu_{UV}= 0.8$ GeV \\
\hline
$2\,L_6-L_4$ &    $ - 0.1 \times  10^{-3}$  \\
$2\,L_8-L_5$ &    $ + 0.4 \times   10^{-3}$ \\
$3\,L_7+L_8$ &    $ - 0.3 \times   10^{-3}$ \\
$m_{\phantom{u}}\,B_0$ & 9.91$ \times  10^{-3}$\,GeV$^2$\\
$m_s\,B_0$& 237$\times   10^{-3}$\,GeV$^2$ \\
$f$  & 92.4 MeV\\
\hline
\end{tabular}
\end{center}
\caption{Low-energy coupling constants used in this work. }
\label{table:Q4_meson_couplings}
\end{table}

The mesonic tadpole $\bar I_Q$  has a logarithmic dependence on the ultraviolet renormalization
scale $\mu_{\rm UV}$. The ultraviolet scale dependence of Eqs. (\ref{result-loop-8}, \ref{result-loop-10}) is counteracted
by the to-be-specified tree-level self-energy of Eq. (\ref{def-Sigma}). Applying a further chiral expansion to Eq. (\ref{def-Sigma}), it
was demonstrated that the physical masses are renormalization-scale-independent. Here, we generalized that result to order N$^3$LO.

In the loop expressions (\ref{result-loop-8}, \ref{result-loop-10}), we use the meson masses accurate to the NNLO as
derived in Refs. \cite{Gasser1985,Bijnens1994}. We recall
\allowdisplaybreaks[1]
\begin{eqnarray}
&& m_\pi^2 =\frac{2\,B_0\,m}{f^2}\,\Big\{ f^2+  \frac{1}{2}\,\bar I_\pi -\frac{1}{6}\,\bar I_\eta
+16\,B_0\,\Big[ (2\, m+m_s)\,(2\,L_6-L_4)+  m\, (2\,L_8-L_5)\Big] \Big\}\,,
\nonumber\\
&& m_K^2 = \frac{B_0\,(m+m_s)}{f^2}\,\Big\{f^2 +
\frac{1}{3}\,\bar I_\eta
\nonumber\\
&& \qquad \qquad +\,16 \,B_0\,\Big[(2\,m+m_s)\,(2\,L_6-L_4)+\frac{1}{2}\,( m+m_s)\,(2\,L_8-L_5)\Big]
\Big\}\,,
\nonumber\\
&& m_\eta^2 = \frac{2\,B_0\,(m+2\,m_s)}{3\,f^2}\,
\Big\{ f^2 + \bar I_K- \frac{2}{3}\, \bar I_\eta
\nonumber\\
&& \qquad \qquad +\, 16\,B_0\,\Big[
(2\, m+m_s)\,(2\,L_6-L_4) +  \frac{1}{3}\, (m + 2\,m_s)\, (2\,L_8-L_5) \Big]
\Big\}
\nonumber\\
&& \quad \;\;\, + \,\frac{2\,B_0\,m}{f^2}\,\left[\frac{1}{6}\, \bar I_\eta-\frac{1}{2}\, \bar I_\pi +\frac{1}{3} \,\bar I_K\right]
+\frac{128}{9}\,\frac{B^2_0\,(m-m_s)^2}{f^2}(3\,L_7+L_8)\,,
\label{meson-masses-q4}
\end{eqnarray}
with the renormalized mesonic tadpole integrals $\bar I_\pi, \bar I_K, \bar I_\eta $ as given in Eq. (\ref{def-tadpole}).
The empirical values for the three relevant combinations $2\,L_6-L_4, 2\,L_8-L_5,3\,L_7+L_8$
are given in Table~\ref{table:Q4_meson_couplings} at the renormalization
scale $\mu_{UV} = 800$ MeV.

We provide a more specific discussion of the renormalized baryon mass parameters that enter the mass
equation (\ref{def-non-linear-system}). With $\bar M_{[8]} $ and $\bar M_{[10]}$, we denote the renormalized form of the bare
parameters  $M_{[8]} $ and $M_{[10]} $. They do not coincide with the chiral SU(3) limit of the baryon masses. The latter are
determined by a set of nonlinear and coupled equations:
\begin{eqnarray}
&& M\ = \bar M_{[8]}   -\, \frac{5 \,C^2  }{768 \,\pi^2\, f^2}\,\frac{\Delta^3\,(2\,M+\Delta)^3}{M^2\,(M+\Delta)^2}\,
\Bigg\{M+ \Delta
\nonumber\\
&& \qquad \qquad +\, \frac{2\,M\,(M+\Delta)+ \Delta^2}{2\,M}  \Bigg\} \,\ln \frac{\Delta^2\,(2\,M+ \Delta )^2}{(M+\Delta)^4} \,,
\nonumber\\
&& M + \Delta =  \bar M_{[10]} -\, \frac{C^2  }{384 \,\pi^2\, f^2}\,\frac{\Delta^3\,(2\,M+\Delta)^3}{(M+\Delta)^4}\,
\Bigg\{M
\nonumber\\
&& \qquad \qquad +\,  \frac{2\,M\,(M+\Delta)+ \Delta^2}{2\,(M+\Delta)}  \Bigg\} \,\ln \frac{M^4}{\Delta^2\,(2\,M+\Delta)^2}\,,
\label{decompose-bare-masses}
\end{eqnarray}
where we identified the baryon octet and decuplet masses in the one-loop self-energy with  $M$ and $M+ \Delta$, respectively. The
baryon  masses receive contributions from the renormalized tree-level terms (\ref{def-non-linear-system}) and from the renormalized
one-loop self-energies (\ref{result-loop-8}, \ref{result-loop-10}). The sum of both provides the scale-invariant chiral limit
values of the octet and decuplet masses, $M$ and $M + \Delta$. Given any values for  $\bar M_{[8]} $ and $\bar M_{[10]} $, the
chiral mass parameters $M$ and $\Delta$ are obtained by a numerical solution of Eq. (\ref{decompose-bare-masses}).  Though in the
chiral limit the self-consistency condition (\ref{def-non-linear-system}) is not a significant effect, a perturbative expansion
of Eq. (\ref{decompose-bare-masses}) in $\Delta/M $ is rapidly converging, this changes as we turn on flavor-breaking effects and increase the quark masses.

We emphasize again that Eqs. (\ref{result-loop-8}, \ref{result-loop-10}) depend on the physical meson and baryon masses
$m_Q$ and $M_{R}$. This defines a self-consistent summation since the masses of the intermediate baryon states
in Eqs. (\ref{result-loop-8}) and (\ref{result-loop-10}) should match the total masses. The baryon masses are a solution of a set
of eight coupled and nonlinear equations in the present scheme. This is a consequence of self-consistency imposed on the
partial summation approach. The latter is a crucial requirement since the loop functions depend sensitively on the precise
values of the baryon masses.

We affirm that a strict chiral expansion  of the expressions
(\ref{result-loop-8}, \ref{result-loop-10}, \ref{result-counter-terms-octet}, \ref{result-counter-terms-decuplet})
to N$^3$LO leads to results that are renormalization scale-independent.

\section{Quark-mass dependence of the baryon masses}

We discuss the determination of the parameter set. In this work, we introduced altogether 41 parameters.
At LO there are 2 parameters $\bar M_{[8]}$ and $\bar M_{[10]}$. At next-to-leading order, there are the 5
parameters $\bar b_{0,D,F}$ and $\bar d_{0,D}$, together with $f= 92.4$ MeV. The parameters $F,D,C$, and $H$ turn
relevant at NNLO. We use the large-$N_c$ sum rules (\ref{large-Nc-HC}) together with $F = 0.45$ and $D =0.8$.
At N$^3$LO, there are 12+5 symmetry-breaking parameters $\bar c_{0-6}, \bar e_{0-4}$ and $\bar \zeta_{0,D,F}, \bar \xi_{0,D}$
introduced in Eq. (\ref{def-chi-34}) and the 17 symmetry-conserving  parameters of Eq. (\ref{def-Q2-terms}).
As discussed in detail in Sec. II, some of the symmetry-breaking parameters are redundant at N$^3$LO.
At N$^3$LO,  there are alltogether 36 relevant parameters. Using large-$N_c$ relations for the N$^3$LO  parameters, the number of
parameters was reduced significantly down to 20. A parameter reduction by about a factor of two was achieved. Still, 20 is a large
number and, in this work, we will consider a subset of the most important operators only. In this work, we will ignore the role of
the five symmetry-conserving N$^3$LO parameters. This leaves us with the 15 parameters, which we adjust to the physical baryon
masses and the quark-mass dependence for the nucleon and omega mass as predicted by the BMW Collaboration \cite{BMW2008}. With
even more accurate and complete QCD lattice results it should be possible to determine the full set of large-$N_c$ correlated
parameters in the near future. The results from the BMW Collaboration are shown in Fig.~2 for three different lattice spacings.
The approximate independence of  the lattice spacing we take as a justification to adjust our parameters without any further
continuum limit extrapolations.

\begin{figure}[t]
\centering
\includegraphics[width=12cm,clip=true]{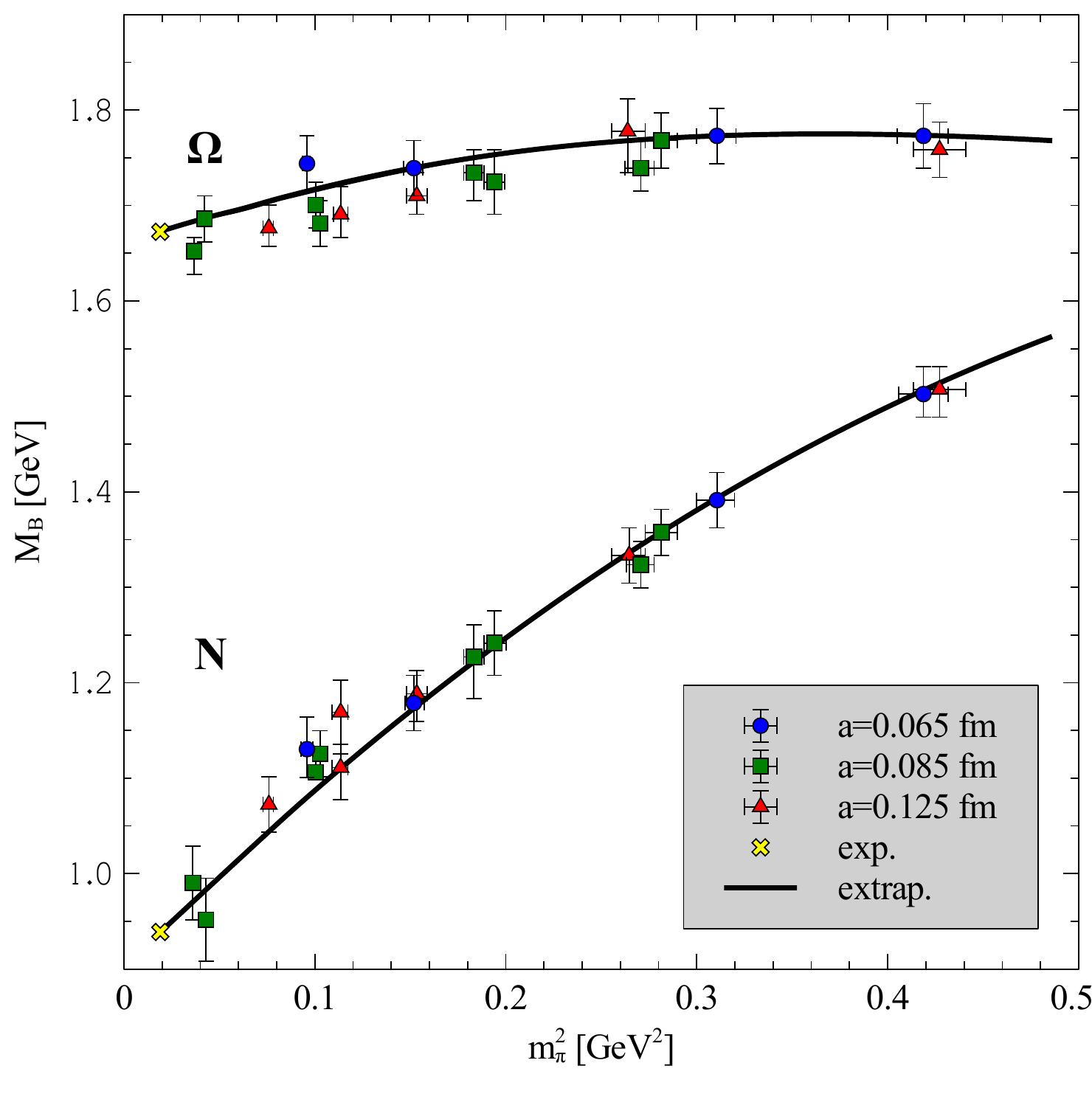}
\caption{Chiral extrapolation of the nucleon and omega masses. Lattice data are taken from Ref. \cite{BMW2008}.}
\label{figure:nucleon_omega_masses_extrapolation_BMW}
\end{figure}

Given a set of parameters there is no guarantee for a unique solution of Eq. (\ref{def-non-linear-system}) to exist. In particular,
there may be a discontinuous quark-mass dependence for the baryon masses. This is a consequence of the nonlinearities in our
approach as introduced by the self-consistency condition. Indeed, various discontinuities in the quark-mass dependence of the
baryon masses were reported in Ref. \cite{Semke2007} based on the chiral Lagrangian relevant at NNLO. While at NNLO, it is not
possible to avoid such a discontinuous quark-mass dependence in our approach; we find that at N$^3$LO, there are parameter sets
that do lead to a smooth quark-mass dependence of the baryon masses.  We observe a necessary condition for a smooth extrapolation:
\begin{eqnarray}
\frac{\partial }{\partial \pslash} \,\Sigma_B (\pslash) \Big|_{\pslash = M_B} < 1 \,,
\label{def-condition}
\end{eqnarray}
to hold for all octet and decuplet self-energies. Owing to our self-consistency constraint, the condition (\ref{def-condition})
depends on the
physical baryon masses and the parameters $\bar \zeta_{0,D,F}, \bar \xi_{0,D}$ only. In order to analyze the
condition (\ref{def-condition})
in more depths, we consider the three mass combinations,
\begin{eqnarray}
&& \Delta_1 ={\textstyle {3\over 4}}\,M_\Lambda + {\textstyle {1\over 4}}\,M_\Sigma
- {\textstyle {1\over 2}}\, \left(M_N - M_\Xi \right)
-{\textstyle {1\over 4}}\, \left( M_{\Sigma^*} - M_\Delta - M_\Omega + M_{\Xi^*} \right)\,,
\nonumber \\
&& \Delta_2= M_\Omega - M_{\Xi^*} - 2\,(M_{\Xi^*}-M_{\Sigma^*}) + M_{\Sigma^*}-M_\Delta \, ,
\nonumber \\
&& \Delta_3= M_{\Sigma^*} - M_\Sigma - M_{\Xi^*} + M_\Xi\,,
\label{def-mass-combinations}
\end{eqnarray}
studied before in Refs. \cite{Dashen1995,Jenkins1995}. As shown in Ref. \cite{Jenkins1995}, a strict large-$N_c$ expansion of
the baryon masses at NNLO predicts $\Delta_1=\Delta_2=\Delta_3=0$. The merit of $\Delta_1$ and $\Delta_2$ lies in their
independence of all parameters but $\bar \zeta_{0,D,F}, \bar \xi_{0,D}$, and $\bar M_{[8]}$. The mass combination $\Delta_3$ has
only an additional dependence on $\bar d_D-3\,(\bar b_F+\bar b_D) $. These properties are a consequence of the self-consistency
constraint and the large-$N_c$ sum rules (\ref{result:large-Nc-chi}), which we use for the renormalized coupling constants at the
renormalization scale $\mu_{UV} = \bar M_{[8]}$.

\begin{table}[t]
\setlength{\tabcolsep}{2.5mm}
\renewcommand{\arraystretch}{1.1}
\begin{center}
\begin{tabular}{l||r|r|r||r|r|r}\hline
 & Fit 1 &  Fit 2  & Fit 3 &  Fit 4$^*$  & Fit 5 &  Fit 6 \\ \hline \hline


$\bar M_{[8]}$\;\, [GeV] &                         0.9138 &               0.9178 &                  0.9203 &                  0.9111 &                0.9159 &                 0.9186  \\
$\bar M_{[10]}$ [GeV]&                         1.0937 &               1.0915 &                  1.0896 &                  1.0938 &                1.0917 &                 1.0897  \\
$\bar b_0\,  \mathrm{[GeV^{-1}]}$ &      -0.9115 &              -0.9234 &                 -0.9355 &                 -0.9086 &               -0.9201 &                -0.9321 \\                    $\bar b_D\,  \mathrm{[GeV^{-1}]}$ &       0.5729 &               0.5978 &                  0.6205 &                  0.5674 &                0.5919 &                 0.6141 \\                    $\bar b_F\,  \mathrm{[GeV^{-1}]}$ &      -0.6322 &              -0.6542 &                 -0.6733 &                 -0.5880 &               -0.6100 &                -0.6289 \\                   $\bar d_0\,  \mathrm{[GeV^{-1}]}$ &      -0.2246 &              -0.2337 &                 -0.2432 &                 -0.2300 &               -0.2385 &                -0.2478 \\                   $\bar d_D\,  \mathrm{[GeV^{-1}]}$ &      -0.3778 &              -0.3691 &                 -0.3586 &                 -0.3617 &               -0.3541 &                -0.3446 \\                   $\bar c_0\,  \mathrm{[GeV^{-3}]}$ &       0.0177 &               0.0022 &                 -0.0135 &                  0.0176 &                0.0020 &                -0.0136 \\                   $\bar c_4\,  \mathrm{[GeV^{-3}]}$ &      -0.1035 &              -0.2269 &                 -0.3413 &                 -0.1659 &               -0.2861 &                -0.3975 \\
$\bar c_5\,  \mathrm{[GeV^{-3}]}$ &      -0.3683 &              -0.3173 &                 -0.2776 &                 -0.3320 &               -0.2824 &                -0.2444 \\
$\bar c_6\,  \mathrm{[GeV^{-3}]}$ &      -1.5436 &              -1.5455 &                 -1.5496 &                 -1.2366 &               -1.2417 &                -1.2482 \\                   $\bar e_4\,  \mathrm{[GeV^{-3}]}$ &      -0.2783 &              -0.2847 &                 -0.2890 &                 -0.2520 &               -0.2605 &                -0.2662 \\                    $\bar \zeta_0\,  \mathrm{[GeV^{-2}]}$ &                          1.7373 &               1.9201 &                  2.0962 &                  1.1279 &                1.3120 &                 1.4885 \\                                    $\bar \zeta_D\,  \mathrm{[GeV^{-2}]}$ &                          0.2582 &               0.2408 &                  0.2244 &                  0.2848 &                0.2670 &                 0.2505 \\                                     $\bar \zeta_F\,  \mathrm{[GeV^{-2}]}$ &                         -0.1949 &              -0.2221 &                 -0.2504 &                 -0.2221 &               -0.2487 &                -0.2769 \\
$\bar \xi_0\,  \mathrm{[GeV^{-2}]}$   &                          1.1999 &               1.4051 &                  1.6083 &                  1.1964 &                1.4026 &                 1.6061 \\

\hline
\end{tabular}
\caption{The parameters are adjusted to reproduce the empirical values of the baryon octet and decuplet masses and the lattice results for quark-mass dependence of  the nucleon and omega as shown in Fig.~1.
The different parameter sets follow with $ \Delta \bar d_D =-0.2 $ GeV$^{-1}$ and $ \Delta \bar d_D =-0.3 $ GeV$^{-1}$ for the first three and last three fits,
respectively. The parameter $\Delta \bar \xi_0$ takes the increasing values $0.2,0.4$ and $0.6 $ in both cases.
}
\label{tab:parameter}
\end{center}
\end{table}

\begin{figure}[t]
\centering
\includegraphics[width=14cm,clip=true]{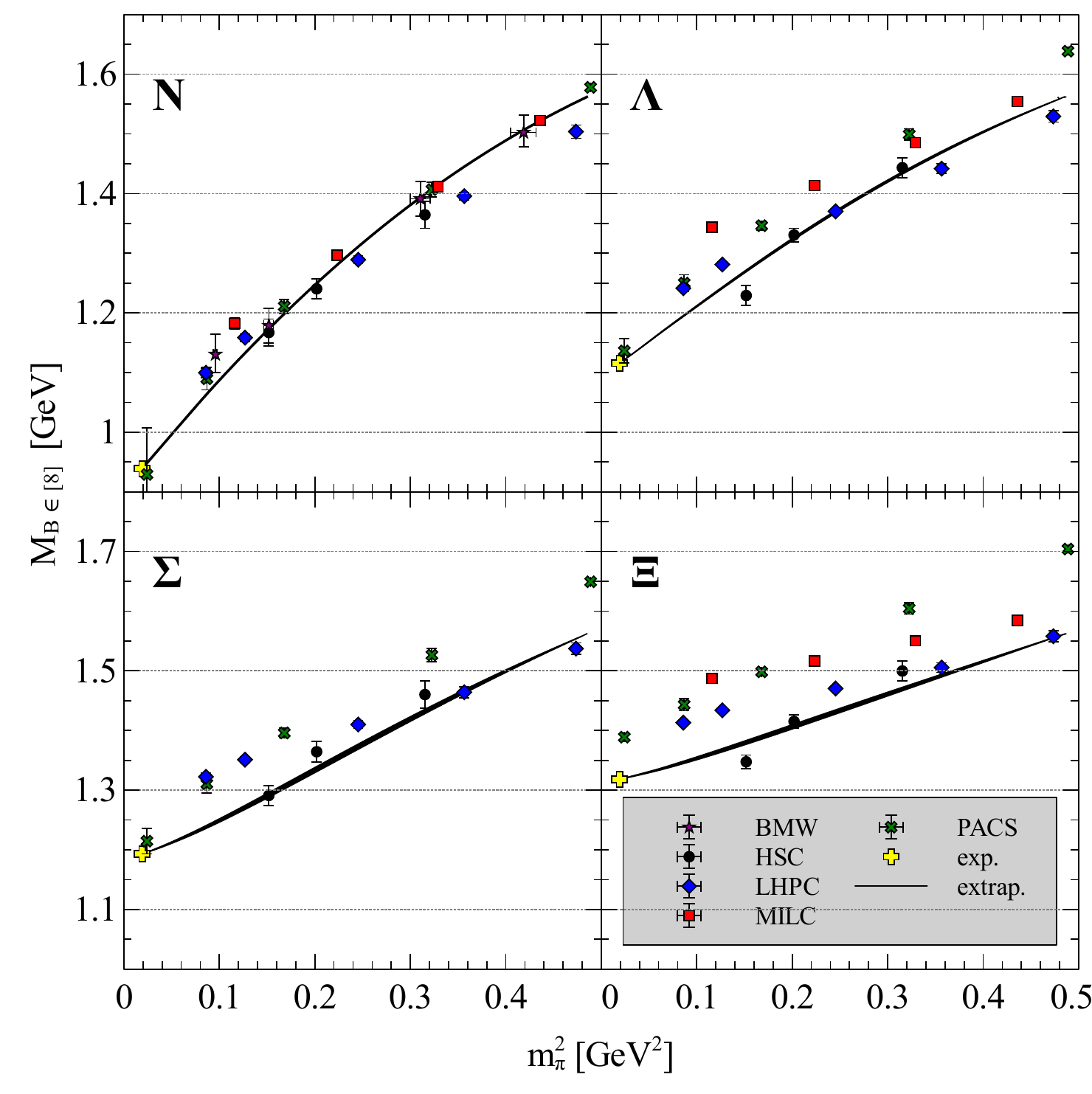}
\caption{Pion-mass extrapolation of the baryon octet masses.}
\label{figure:lattice_data_baryon_extrapolation-octet}
\end{figure}

\begin{figure}[t]
\centering
\includegraphics[width=14cm,clip=true]{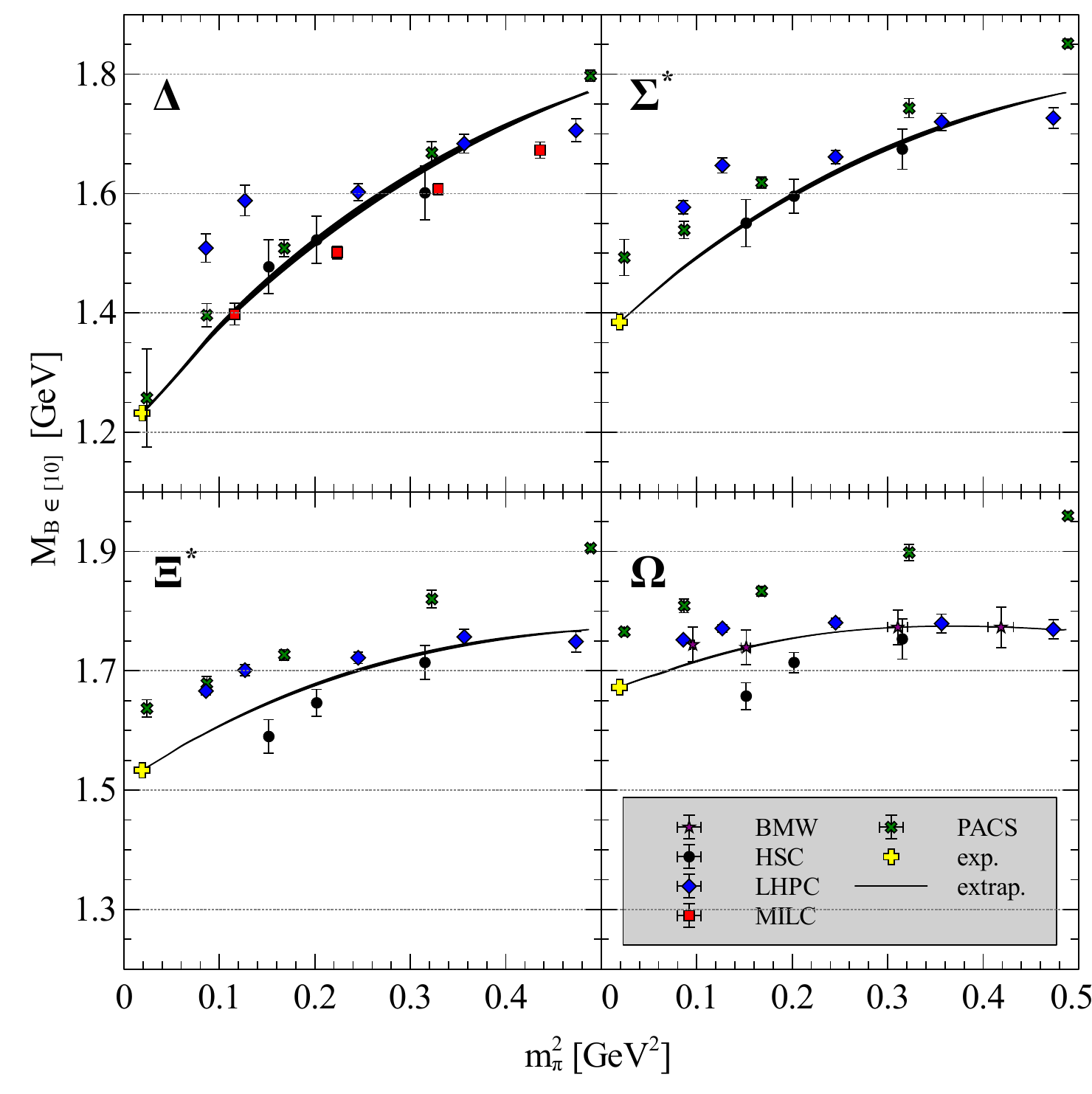}
\caption{Pion-mass extrapolation of the baryon decuplet masses.}
\label{figure:lattice_data_baryon_extrapolation-decuplet}
\end{figure}

Using the empirical values for $\Delta_1 \simeq 3.2$ MeV and $\Delta_2\simeq -6.1$ MeV, together with the large-$N_c$ sum rule
$\bar \xi_D=3\,(\bar \zeta_F+\bar \zeta_D) $ from Eq. (\ref{result:large-Nc-zeta}), we can determine the parameters
$\bar \zeta_{F,D}$ in terms of the two singlet parameters $\bar \zeta_0$ and $\bar \xi_0$. In turn, the
condition (\ref{def-condition}) may be considered as a function of those parameters
$\bar \zeta_0$ and $\bar \xi_0$ only. We find
\begin{eqnarray}
\bar \xi_0 >  \bar \xi^{\rm crit}_0 \simeq  1.1070 +1.1882\,\ln \bar M_{[8]}  \qquad {\rm  for} \qquad -1.5<\bar \zeta_0 <1.5 \,,
\end{eqnarray}
with $\bar M_{[8]}$ measured in units of GeV.
The third mass combination with its empirical value $\Delta_3 \simeq -23.9$ MeV leads to the condition
\begin{eqnarray}
&&\bar \zeta_0 =  3.1398  + 2.5604\,\ln \bar M_{[8] }
+0.5800\,\Delta \bar \xi_0 +\Delta d_D\,\Big( 6.2490 +2.3801 \ln \bar M_{[8] }  \Big)
\nonumber\\
&& \quad \, -\,\Big( \bar M_{[10]}-\bar M_{[8]}\Big)\,\Big(  0.9017 +  1.4407 \,\ln \bar M_{[8] }
-1.6120\,\Delta \bar \xi_0 \Big) \,,
\nonumber\\
&& \bar \xi_0 = \bar \xi_0^{\rm crit}+ \Delta \bar \xi_0\,, \qquad \qquad
\Delta \bar d_D = \bar d_D - 3\,\bar b_F-3\,\bar b_D\,,
\label{zeta0-correlation}
\end{eqnarray}
where  all parameters are assumed in units of GeV. From the large-$N_c$ sum rule (\ref{largeNcbd}) we expect $\Delta \bar d_D =0$.
As seen from (\ref{zeta0-correlation}) this would lead to unnaturally large values for $\bar \zeta_0$, at least for reasonable
choices of  $ \bar M_{[8] }$ and $ \bar M_{[10] }$. Since the parameter $\bar d_D$ enters at NLO it is justified
to admit a small $\Delta \bar d_D < 0$, as it would arise at the next order in the large-$N_c$ expansion.

In the following, we assume fixed values for $\Delta \bar d_D $ and  $\Delta \bar \xi_0 $ and adjust the remaining 14 parameters
to the physical baryon masses and the pion-mass dependence of the nucleon and omega masses as predicted by the BMW Collaboration.
The results are shown in Fig.~1 for various choices. We find that the pion-mass dependence of the BMW results can be reproduced
accurately for any given $\Delta \bar d_D $ and  $\Delta \bar \xi_0 $. The size of the fitted parameters are collected in
Table \ref{tab:parameter}. While the parameters $\Delta \bar d_D$ and $\Delta \bar \xi_0$ cannot be determined from the BMW results, the request for natural-size parameters favors Fit 4, for which all parameters take a reasonable size. The chiral limit
values of the baryon octet and decuplet states, $M$ and $M+ \Delta$, follow from the solution of the set of  nonlinear
equations (\ref{decompose-bare-masses}). Using the parameters of Table \ref{tab:parameter}, we find the ranges
\begin{eqnarray}
M = 943.9 \pm 1.8 \,{\rm MeV} \,, \qquad \quad
M+ \Delta  = 1085.8 \pm 1.6\, {\rm MeV} \,.
\end{eqnarray}

In Fig.~2 and 3 we confront our results for the baryon octet and decuplet masses with the predictions of various lattice
groups. The almost invisible bands in the figure are generated by the 6 parameter sets as specified in Table \ref{tab:parameter}.
We refrain from incorporating any finite lattice effects in our present study, so the comparison in Figs.~2 and 3, is in part, of a qualitative nature. The spread in the various lattice simulation results may be taken as an indication on the
size of different finite lattice effects. Most interesting is the comparison of our results with the predictions from
HSC \cite{HSC2008}, for which one may expect the need of only minor lattice corrections. We find it encouraging that our
approach appears to recover the pion-mass dependence of the unfitted baryon masses of HSC \cite{HSC2008} reasonably well.

\section{Summary}

We have studied the pion-mass dependence of the baryon octet and decuplet masses based on the chiral Lagrangian truncated at
N$^3$LO. The large number of parameters was reduced significantly in application of large-$N_c$ sum rules and therewith
allowed for a first meaningful analysis of recent QCD lattice results. Altogether, we considered 16 parameters, where we
ignored the small effects of the 5 symmetry-conserving N$^3$LO parameters. In our analysis, we used a covariant form of the
chiral Lagrangian and the pertinent loop functions relevant at N$^3$LO. Owing to a self-consistency condition, which requires
the use of physical masses in the one-loop functions, a successful reproduction of the recent results of the
BMW Collaboration on the nucleon and omega mass was achieved. A smooth quark-mass dependence arose upon a suitable choice of
the symmetry-breaking N$^3$LO parameters. A prediction for the pion-mass dependence of the remaining octet and decuplet masses
was presented and confronted with available unquenched three-flavor simulations of various lattice groups.
We recover the recent results of the HSC without any further adjustments.

With additional lattice data, in particular, on the dependence of the baryon masses on the strange quark mass, it should be
possible to determine the remaining five symmetry-conserving parameters and scrutinize in more depths the reliability of the
assumed large-$N_c$ sum rules.

\clearpage

\begin{appendix}

\section{Tree-level baryon self-energy}

We specify the renormalized tree-level self-energies for the baryon octet and decuplet states.
There are several contributions to $\Sigma^{\rm tree-level}$ in our scheme. We express the tree-level self-energy
in terms of the renormalized coupling constants. It holds:
\allowdisplaybreaks[1]
\begin{eqnarray}
\Sigma^{\rm tree-level}_N &=& - 4\,B_0 \left( \bar b^{\rm eff}_0\, (2\,m+m_s) + \bar b^{\rm eff}_D\, (m+m_s) + \bar b^{\rm eff}_F\, (m-m_s) \right)
\nonumber \\
&\phantom =& -4\,B_0^2\, \Big( \bar c_0\, (2\,m^2 + m_s^2) + \bar c_2\, (m^2+m_s^2) + \bar c_3\, (m^2-m_s^2) \Big)
\nonumber \\
&\phantom =&  - 2\,B_0\, \Big( \bar \zeta_0\, (2\,m+m_s) + \bar \zeta_D\, (m+m_s)
+ \bar \zeta_F\, (m-m_s) \Big) \Big(M_N - \bar M_{[8]} \Big) \,,
\nonumber \\
\Sigma^{\rm tree-level}_\Lambda &=& - 4\,B_0\, \Big(\bar b^{\rm eff}_0\, (2\,m+m_s) + \frac{2}{3}\,\bar b^{\rm eff}_D\, (m+2\,m_s) \Big)
\nonumber \\
&\phantom =& - 4\,B_0^2\, \Big( \bar c_0\, (2\,m^2 + m_s^2) + \frac 23 \,\bar c_1\, (m - m_s)^2  + \frac 23 \,\bar c_2\, (m^2 + 2\,m_s^2) \Big)
\nonumber \\
&\phantom =& - 2\,B_0\, \Big(\bar \zeta_0\, (2\,m+m_s) + \frac{2}{3}\,\bar \zeta_D\, (m+2\,m_s) \Big)  \Big(M_\Lambda - \bar M_{[8]} \Big)\,,
\nonumber \\
\Sigma^{\rm tree-level}_\Sigma&=& - 4\,B_0\, \Big(\bar b^{\rm eff}_0\, (2\,m+m_s) + 2\,\bar b^{\rm eff}_D\,m \Big)
\nonumber \\
&\phantom =&  - 4\,B_0^2\, \Big(\bar c_0\, (2\,m^2+m_s^2) + 2\,\bar c_2\, m^2 \Big)
\nonumber \\
&\phantom =& - 2\,B_0\, \Big(\bar \zeta_0\, (2\,m+m_s) + 2\,\bar \zeta_D\,m
\Big)  \Big(M_\Sigma - \bar M_{[8]} \Big) \,,
\nonumber \\
\Sigma^{\rm tree-level}_\Xi &=& - 4\,B_0\, \Big( \bar b^{\rm eff}_0\, (2\,m+m_s)  + \bar b^{\rm eff}_D\, (m+m_s)  - \bar b^{\rm eff}_F\, (m - m_s) \Big)
\nonumber\\
&\phantom =& - 4\,B_0^2\left( \bar c_0\, (2\,m^2+m_s^2) + \bar c_2\,(m^2 + m_s^2) - \bar c_3\,(m^2 - m_s^2) \right)
\nonumber\\
&\phantom =& - 2\,B_0\, \Big( \bar \zeta_0\, (2\,m+m_s)  + \bar \zeta_D\,
(m+m_s)  - \bar \zeta_F\, (m - m_s) \Big)  \Big(M_\Xi - \bar M_{[8]} \Big) \,,
\label{result-counter-terms-octet}
\end{eqnarray}
and
\allowdisplaybreaks[1]
\begin{eqnarray}
\Sigma^{\rm tree-level}_\Delta &=& - 4\,B_0\, \Big( \bar d^{\rm eff}_0\, (2\,m+m_s)  + \bar d^{\rm eff}_D\, m\Big)
 \nonumber\\
&\phantom =&  - 4\,B_0^2\, \Big( \bar e_0\,(2\,m^2 + m_s^2) + \bar e_2\, m^2 \Big)
 \nonumber\\
&\phantom =&  - 2\,B_0\, \Big( \bar \xi_0\, (2\,m+m_s)  +
\bar \xi_D\, m\Big)  \Big(M_{\Delta } - \bar M_{[10]} \Big) \,,
\nonumber \\
\Sigma^{\rm tree-level}_{\Sigma^*} &=& - 4\,B_0\, \Big( \bar d^{\rm eff}_0\, (2\,m+m_s)  + \frac{1}{3} \,\bar d^{\rm eff}_D\, (2\,m+m_s) \Big)
\nonumber \\
&\phantom =& - 4\,B_0^2\, \Big( \bar e_0\, (2 \,m^2 + m_s^2) + \frac 13 \,\bar e_1\, (m - m_s)^2 + \frac 13\, \bar e_2\,(2\,m^2 + m_s^2) \Big)
\nonumber \\
&\phantom =& - 2\,B_0\, \Big( \bar \xi_0\,
(2\,m+m_s)  + \frac{1}{3}\, \bar \xi_D\, (2\,m+m_s) \Big)
\Big(M_{\Sigma^*} - \bar M_{[10]} \Big)\,,
\nonumber \\
\Sigma^{\rm tree-level}_{\Xi^*} &=& - 4\,B_0\, \Big( \bar d^{\rm eff}_0\, (2\,m+m_s)  + \frac{1}{3} \,\bar d^{\rm eff}_D\, (m+2\,m_s) \Big)
\nonumber \\
&\phantom =& -4\,B_0^2\, \Big( \bar e_0\, (2\,m^2 + m_s^2)  + \frac 13\, \bar e_1\, (m - m_s)^2 + \frac 13\, \bar e_2\, (m^2 + 2\,m_s^2) \Big)\,,
\nonumber \\
&\phantom =& - 2\,B_0\, \Big( \bar \xi_0\, (2\,m+m_s)
+ \frac{1}{3} \,\bar \xi_D\, (m+2\,m_s) \Big)  \Big(M_{\Xi^*} - \bar M_{[10]} \Big)\,,
\nonumber \\
\Sigma^{\rm tree-level}_\Omega &=& - 4\,B_0\, \Big( \bar d^{\rm eff}_0\, (2\,m+m_s) + \bar d^{\rm eff}_D\, m_s \Big)
\nonumber \\
&\phantom =&  - 4\, B_0\, \Big( \bar e_0\, (2\,m^2+m_s^2) + \bar e_2\, m_s^2 \Big)
\nonumber \\
&\phantom =&   - 2\,B_0\, \Big( \bar \xi_0\, (2\,m+m_s) +
\bar \xi_D\, m_s \Big) \,\Big(M_\Omega - \bar M_{[10]} \Big)\,,
\label{result-counter-terms-decuplet}
\end{eqnarray}
with
\begin{eqnarray}
&&\bar b^{\rm eff}_0 \equiv \bar b_0+\bar c_6\, B_0\,(2\,m+m_s)\,, \qquad
 \bar b^{\rm eff}_D \equiv \bar b_D+\bar c_4\, B_0\,(2\,m+m_s)\,, \qquad
\nonumber\\
&&  \bar b^{\rm eff}_F \equiv \bar b_F+\bar c_5\, B_0\,(2\,m+m_s)\,,
\nonumber\\
&& \bar d^{\rm eff}_0 \equiv \bar d_0+\bar e_4 \,B_0\,(2\,m+m_s),\qquad \bar d^{\rm eff}_D \equiv \bar d_D+\bar e_3\, B_0\,(2\,m+m_s)\,.
\end{eqnarray}

\end{appendix}

\newpage
\bibliography{literatur}
\end{document}